\documentclass[journal=ancac3,manuscript=article]{achemso}

\usepackage[version=3]{mhchem} 

\usepackage{xcolor}

\author{Cheng Ji}
\affiliation[a]{Pritzker School of Molecular Engineering, University of Chicago, Chicago, IL 60637}
\alsoaffiliation[b]{Materials Science Division, Argonne National Laboratory, Lemont, IL 60439}
\author{Michael T. Solomon}
\affiliation[a]{Pritzker School of Molecular Engineering, University of Chicago, Chicago, IL 60637}
\alsoaffiliation[b]{Materials Science Division, Argonne National Laboratory, Lemont, IL 60439}
\alsoaffiliation[c]{Center for Molecular Engineering, Argonne National Laboratory, Lemont, IL 60439}
\author{Gregory D. Grant}
\affiliation[a]{Pritzker School of Molecular Engineering, University of Chicago, Chicago, IL 60637}
\alsoaffiliation[b]{Materials Science Division, Argonne National Laboratory, Lemont, IL 60439}
\author{Koichi Tanaka}
\affiliation[a]{Pritzker School of Molecular Engineering, University of Chicago, Chicago, IL 60637}
\author{Muchuan Hua}
\affiliation[e]{Nanoscience and Technology Division, Argonne National Laboratory, Lemont, IL 60439}
\author{Jianguo Wen}
\affiliation[e]{Nanoscience and Technology Division, Argonne National Laboratory, Lemont, IL 60439}
\author{Sagar K. Seth}
\affiliation[a]{Pritzker School of Molecular Engineering, University of Chicago, Chicago, IL 60637}
\alsoaffiliation[b]{Materials Science Division, Argonne National Laboratory, Lemont, IL 60439}
\author{Connor P. Horn}
\affiliation[a]{Pritzker School of Molecular Engineering, University of Chicago, Chicago, IL 60637}
\alsoaffiliation[b]{Materials Science Division, Argonne National Laboratory, Lemont, IL 60439}
\author{Ignas Masiulionis}
\affiliation[a]{Pritzker School of Molecular Engineering, University of Chicago, Chicago, IL 60637}
\alsoaffiliation[b]{Materials Science Division, Argonne National Laboratory, Lemont, IL 60439}
\author{Manish K. Singh}
\affiliation[a]{Pritzker School of Molecular Engineering, University of Chicago, Chicago, IL 60637}
\altaffiliation{Current address: memQ Inc., Chicago, IL 60615}
\author{Sean E. Sullivan}
\affiliation[b]{Materials Science Division, Argonne National Laboratory, Lemont, IL 60439}
\alsoaffiliation[c]{Center for Molecular Engineering, Argonne National Laboratory, Lemont, IL 60439}
\altaffiliation{Current address: memQ Inc., Chicago, IL 60615}
\author{F. Joseph Heremans}
\affiliation[b]{Materials Science Division, Argonne National Laboratory, Lemont, IL 60439}
\alsoaffiliation[c]{Center for Molecular Engineering, Argonne National Laboratory, Lemont, IL 60439}
\alsoaffiliation[a]{Pritzker School of Molecular Engineering, University of Chicago, Chicago, IL 60637}
\author{David D. Awschalom}
\affiliation[a]{Pritzker School of Molecular Engineering, University of Chicago, Chicago, IL 60637}
\alsoaffiliation[b]{Materials Science Division, Argonne National Laboratory, Lemont, IL 60439}
\alsoaffiliation[c]{Center for Molecular Engineering, Argonne National Laboratory, Lemont, IL 60439}
\alsoaffiliation[d]{Department of Physics, University of Chicago, Chicago, IL 60637}
\author{Supratik Guha}
\affiliation[a]{Pritzker School of Molecular Engineering, University of Chicago, Chicago, IL 60637}
\alsoaffiliation[b]{Materials Science Division, Argonne National Laboratory, Lemont, IL 60439}
\alsoaffiliation[c]{Center for Molecular Engineering, Argonne National Laboratory, Lemont, IL 60439}
\email{sguha@anl.gov}
\author{Alan M. Dibos}
\affiliation[e]{Nanoscience and Technology Division, Argonne National Laboratory, Lemont, IL 60439}
\alsoaffiliation[c]{Center for Molecular Engineering, Argonne National Laboratory, Lemont, IL 60439}
\email{adibos@anl.gov}

\title[An \textsf{achemso} demo]
  {Nanocavity-mediated Purcell enhancement of Er in \ce{TiO2} thin films grown via atomic layer deposition}

\abbreviations{}

\begin{document}







\begin{abstract}
The use of trivalent erbium (\ce{Er^3+}), typically embedded as an atomic defect in the solid-state, has widespread adoption as a dopant in telecommunications devices and shows promise as a spin-based quantum memory for quantum communication. In particular, its natural telecom C-band optical transition and spin-photon interface makes it an ideal candidate for integration into existing optical fiber networks without the need for quantum frequency conversion. However, successful scaling requires a host material with few intrinsic nuclear spins, compatibility with semiconductor foundry processes, and straightforward integration with silicon photonics. Here, we present Er-doped titanium dioxide (\ce{TiO2}) thin film growth on silicon substrates using a foundry-scalable atomic layer deposition process with a wide range of doping control over the Er concentration. Even though the as-grown films are amorphous, after oxygen annealing they exhibit relatively large crystalline grains, and the embedded Er ions exhibit the characteristic optical emission spectrum from anatase \ce{TiO2}. Critically, this growth and annealing process maintains the low surface roughness required for nanophotonic integration. Finally, we interface Er ensembles with high quality factor Si nanophotonic cavities via evanescent coupling and demonstrate a large Purcell enhancement ($\sim300$) of their optical lifetime. Our findings demonstrate a low-temperature, non-destructive, and substrate-independent process for integrating Er-doped materials with silicon photonics. At high doping densities this platform can enable integrated photonic components such as on-chip amplifiers and lasers, while dilute concentrations can realize single ion quantum memories.

\end{abstract}

\newpage

Future quantum networks may bring tremendous improvement to realms of secure communication, distributed quantum computing, and remote quantum sensing.\cite{Wehner2018} In order for quantum networks to achieve reliable information transfer over long distances with suitable bandwidth, the network nodes must eventually employ quantum repeaters to overcome lossy single photon channels.\cite{Briegel1998, Lvovsky2009, Sangouard2011, Singh2021} Trivalent erbium (\ce{Er^3+} or Er for brevity) is a workhorse emitter used in traditional telecommunications technology, and Er ions have recently emerged as promising communication qubits due to their narrow telecom C-band optical transition\cite{Thomas, Serrano2022} and long electron spin coherence times.\cite{Zhong2015, Wang2017, Rancic2018} Individual Er ions have already shown promise as a quantum memory element in repeaters, demonstrating storage via the long spin state for use in future entanglement swapping protocols.\cite{Raha2020, Chen2020, LeDantec2021} In addition, photonics-integrated ensembles of Er ions have also been applied in microwave-to-optical quantum state transduction,\cite{Barnett2020, Xie2021, Rochman2023} to address the challenge of interfacing the optical photons used in quantum networks ($\sim195$ THz) with processing nodes operating at microwave frequencies ($\sim$10 GHz). However, despite these benefits, the long-lived optical lifetime of the telecom C-band transition has limited its development, as nanophotonic cavities are necessary to enhance light-matter interactions and greatly decrease the photon excited state lifetime via Purcell enhancement. Beyond quantum applications, successful integration of Er with silicon photonics can reduce the device footprint and reduce power requirements for classical telecommunications devices\cite{Wang2022}, such as frequency synthesizers\cite{Xin2019}, modulators\cite{Li2021}, on-chip optical gain media for light sources,\cite{Ronn2016, Ronn2019} and integrated telecom amplifiers.\cite{Chen:22} Thus, regardless of application there is an emerging need for scalable, foundry-compatible growth techniques, which exhibit large dopant tunability, and fabrication processes that integrate these materials into on-chip photonic platforms.

To address the challenges in integrating silicon photonics with functional Er-doped materials for quantum networking nodes, recent studies have demonstrated enhanced emission and control of a single Er ion via flip-chip bonded silicon nanophotonic cavities on the surface of solid-state Er-doped bulk crystals.\cite{Dibos2018, Uysal2023, Ourari2023} However, to be scalable at the foundry-level, it is highly preferrable  to couple Er to on-chip nanophotonic cavities via deposition and patterning of Er-doped films using well-established semiconductor processes. This is most easily achieved by doping Er ions in a suitable host material that is compatible with nanophotonic platforms such as silicon-on-insulator (SOI).\cite{Dibos2022, Gritsch2022, Gritsch2023} Previous studies with SOI based photonics have used ion implantation\cite{Gritsch2022, Gritsch2023} to directly embed Er into devices or molecular beam deposition\cite{Dibos2022} to grow thin film Er-doped crystals on top of SOI followed by fabrication of nanophotonic cavities. While both techniques are effective, they have drawbacks. Ion implantation is relatively straightforward, but the large mass of Er ion can lead to excessive lattice damage upon impact, creating interstitial sites and deleterious decay pathways. The crystalline damage is often difficult to fully repair, even after high temperature annealing. In contrast, molecular beam deposition with in-situ Er doping can produce high-quality thin films, but it is expensive, needs to be performed in ultra-high vacuum, and growth parameters are highly dependent on the substrate. Overall these complexities make foundry-level processing quite challenging and prevent the integration of various active components with photonic circuits at scale.

An alternative approach enabling such integration is to grow high-quality thin films via atomic layer deposition (ALD)\textemdash a low-temperature, non-destructive, and substrate-independent process. ALD offers significant advantages over molecular beam epitaxy as a manufacturing method for a variety of reasons including scalability, throughput, cost of operation, deposition conformality, and thin film uniformity; and as a result, ALD is already ubiquitous in the semiconductor industry.\cite{Biyikli2017, Shahin2018, Rafaiel2019, Bhaswar2021, Zheng2022} In ALD, gaseous chemical precursors react on the substrate surface in a layer-by-layer deposition process, with excess precursors and by-products purged out between each reaction step. This process allows for cyclic layer-by-layer deposition, offering precise control over film thickness and conformality. Studies on Er-doped dielectric oxides grown by ALD have been sparse, though ALD-grown Er-doped \ce{Al2O3} thin films have been used for on-chip nanolasers\cite{Ronn2016} and have shown potential for integration with waveguides.\cite{Ronn2019} However, the reported optical linewidth in amorphous aluminum oxide has been too broad for quantum communication applications. To address this limitation, titanium dioxide (\ce{TiO2}) has been proposed as a desirable host material because it has a low natural abundance of intrinsic nuclear spins minimizing decoherence and a relatively large optical bandgap.\cite{Dette2014,Kanai2021, Stevenson2022} Experimental demonstrations have shown narrow ensemble optical and spin linewidths for erbium when implanted into bulk rutile phase \ce{TiO2}, albeit after high temperature annealing.\cite{Phenicie2019} Reasonably narrow inhomogeneous optical linewidths have also been achieved with molecular beam epitaxy of thin film single crystal anatase \ce{TiO2} on \ce{LaAlO3} substrates.\cite{Shin2022} 

Recently we presented preliminary results on low temperature ALD growth of Er-doped \ce{TiO2} thin films, which were directly deposited on Si and annealed.\cite{CJ2023} Our characterization included the crystalline phase, surface roughness, and inhomogeneous linewidth via x-ray diffraction (XRD), atomic force microscopy (AFM), and photoluminescence excitation (PLE) spectroscopy, respectively. In this article, we expound on these results by investigating the anatase phase \ce{TiO2} crystalline material that emerges after annealing. We determine the large grain size and crystallinity using electron microscopy and characterize the large range of Er doping via secondary ion mass spectrometry (SIMS). With a better understanding of the material quality, we demonstrate the applicability of ALD grown films to nanophotonic devices, by fabricating and measuring 1D photonic crystal cavities using ALD Er:\ce{TiO2} grown on SOI wafers. Our findings suggest that Er:\ce{TiO2} thin films synthesized via ALD and integrated with silicon photonics provide a scalable, foundry-compatible growth technique, which exhibits large dopant tunability able to meet the emerging needs of both classical and quantum technologies.


\section{Results and Discussion}

\subsection{Er-Doped Thin Film \ce{TiO2} Growth}

The Er-doped \ce{TiO2} thin films were grown using a plasma-enhanced ALD system, which can be operated with and without plasma. The deposition process involved the use of titanium isopropoxide (TTIP) and water (\ce{H2O}) as precursors for the thermal deposition of \ce{TiO2} (additional growth details are discussed in the \textit{Methods}).\cite{Niemela2017, Hackler2019, Donnell2022} We chose TTIP as the Ti precursor to minimize impurities in the as-grown \ce{TiO2} thin films because of its composition, which comprises only elements of carbon, hydrogen, titanium, and oxygen.\cite{Kim2022} For the Er doping, we selected Tris(2,2,6,6-tetramethyl-3,5-heptanedionato)erbium (\ce{Er(thd)_3} or \ce{Er(tmhd)_3}), as it can react with oxygen plasma to form atomic layers of erbium oxides, thereby incorporating erbium into various host materials.\cite{Ronn2016, Van2006, Hoang2007, Van2006_1, Van2005} In an effort to grow smooth films, we deliberately maintained the Si substrates at a temperature of 120 $^{\circ}$C during the deposition \textemdash well below the 150-165 $^{\circ}$C temperature range for deposition of anatase phase \ce{TiO2}\cite{Aarik1995}\textemdash at the expense of amorphous growth.\cite{Niemela2017}

\begin{figure}
    \includegraphics[width=0.98\textwidth]{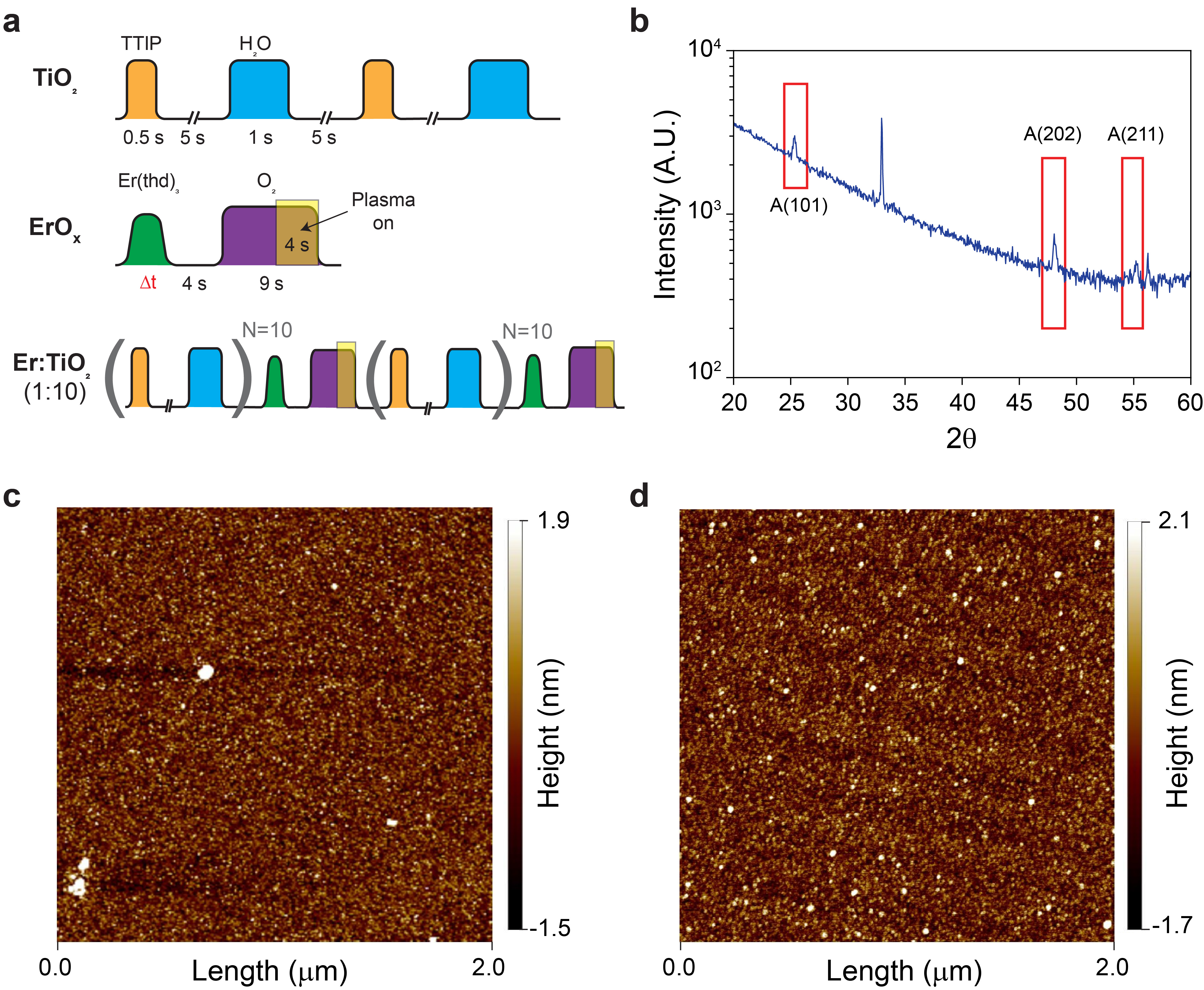}
    \caption{The ALD process and material characterization. (a) ALD deposition process pulse sequence. Top (\ce{TiO2}): Two cycles of thermal ALD growth, where the TTIP precursor and water are successively pulsed into the chamber and reacted to form a \ce{TiO2} monolayer. Middle (\ce{ErO_x}): A single cycle where the \ce{Er(thd)_3} is pulsed into the chamber with a bubbler and a turbo pump is used to remove excess precursor. The residual precursor reacts with oxygen plasma, which is turned on during the last 4 seconds of the oxygen pulse, to create a layer of erbium oxide. The \ce{Er(thd)_3} pulse length ($\Delta t$, marked in red) can be varied to control the doping level. Bottom (Er:\ce{TiO2}): Composite growth, where one \ce{ErO_x} deposition cycle is followed by 10 cycles of \ce{TiO2} deposition to space out the Er ions. The doping concentration can be further controlled by adjusting the \ce{Er(thd)_3} evaporation temperature and pulse length. (b) XRD spectrum of an oxygen annealed Er-doped \ce{TiO2} thin film grown on a Si(100) substrate. The thin film comprises a `$10/10/10$' structure, as described in the main text. The XRD spectrum highlights the characteristic peaks of anatase \ce{TiO2} in the red boxes. (c-d) AFM height scans of (c) as-grown and (d) annealed thin films from the same growth run. The AFM scan area measures $2$ $\mu$m $\times$~2~$\mu$m. The root mean square roughness (Rq) and arithmetic mean roughness (Ra) values for (c) are 0.522 nm and 0.348 nm, respectively. The Ra and Rq values for (d) are 0.520 nm and 0.399 nm, respectively.}
    \label{fig:1}
\end{figure}

The Er:\ce{TiO2} ALD pulse sequence is depicted schematically in Figure \ref{fig:1}a. Similar in overall sequence to that demonstrated for \ce{Er{:}Al2O3},\cite{Ronn2016} we employ an alternating sequence of undoped \ce{TiO2} layers interrupted with single cycles of \ce{ErO_x} to provide the Er doping. For the undoped thermal ALD \ce{TiO2} deposition steps (Fig.~\ref{fig:1}a, top), we pulse the TTIP precursor into the chamber for 0.5 s and the water for 1 s, where the additional \ce{H2O} pulse time is used to fully consume the the TTIP. We use a 5-second window after each pulse to pump out excess precursor, thereby allowing for a more homogeneous deposited layer. For the Er doping process, we introduce \ce{Er(thd)_3} into the chamber via a bubbler where it binds to the substrate (Fig. \ref{fig:1}a, middle). The \ce{Er(thd)_3} pulse length is variable ($\Delta t$), to coarsely control the Er concentration. Following a 4-second wait time, we pulse molcular oxygen into the chamber and, synchronously, we open a turbo molecular pumping line to quickly remove excess precursors. After 5 s of \ce{O2} flow, we ignite the oxygen plasma for 4 s, reacting with the \ce{Er(thd)_3} on the surface to form atomic layers of erbium oxide. Upon plasma cessation, we close valve to the turbo pump, completing one cycle of \ce{ErO_x} deposition. In this study, the growth cycle ratio was fixed at 1 cycle of ErO$_x$ for every 10 cycles of \ce{TiO2}, as delineated in the lower panel of Figure \ref{fig:1}a. After the deposition, we remove the sample from the reactor and perform ex-situ thermal annealing at 400 $^{\circ}$C for 30 min in a pure oxygen environment.

We can make use of two primary methods to controlling the concentration of dopants during ALD growth: (1) altering the \ce{TiO2} deposition ratio relative to \ce{ErO_x} deposition, or (2) controlling the growth rate of the \ce{ErO_x} layer itself. In the first approach, the overall Er concentration is controlled by modifying the \ce{TiO2} spacing (via insertion of more cycles of TTIP/\ce{H2O}) around each \ce{ErO_x} layer. However, this method, while feasible and easy to apply in ALD processes, has a limited tuning range from a few percent to a few tenths of a percent due to the finite number of cycles required to grow a thin film. In this study we focused on the second method of controlling doping by tuning the erbium oxide deposition rate. For example, to decrease the Er concentration, we can lower the \ce{Er(thd)_3} evaporation temperature and shorten the precursor pulse. Consequently, in each cycle, a reduced amount of \ce{Er(thd)_3} will enter the reaction chamber.

\subsection{Morphology and Doping Characterization}

Previous optical experiments have shown that the inclusion of undoped `buffer' and `capping' layers can reduce optical inhomogeneous broadening for epitaxial Er:\ce{Y2O3} films,\cite{Singh2020} epitaxial Er:\ce{TiO2} thin films,\cite{Shin2022} as well as spectral diffusion and inhomogeneous linewidth broadening for Er in polycrystalline \ce{TiO2} films.\cite{singh2022} While a thicker layer is better for optical properties, it is more difficult to etch for eventual nanophotonic devices. With both of these considerations in mind, the films used for our optical studies were grown with a `$10/10/10$' doped heterostructure: a 10 nm undoped \ce{TiO2} buffer layer at the Si interface, a 10 nm Er-doped \ce{TiO2} layer in the middle, and a 10 nm undoped \ce{TiO2} capping layer on top. Figure \ref{fig:1}b displays the XRD spectrum of an oxygen annealed `$10/10/10$' heterostructure thin film sample. The spectrum shows the characteristic peaks of anatase \ce{TiO2} \cite{Horprathum2010} (marked with red boxes), though the signature is weak due to the modest thickness of the film, as well as the peak from the Si substrate (around 33$^{\circ}$).\cite{Zaumseil2015} The oxygen annealing process enhances the polycrystallinity of the as-grown amorphous film, as has been seen previously.\cite{Niilisk2006} To evaluate the surface morphology we measured the roughness of the as-grown and annealed `$10/10/10$' films using AFM, as shown in Figures \ref{fig:1}c and \ref{fig:1}d, respectively. The AFM height maps yield surface roughness figures of merit, Ra and Rq, that are very similar for as-grown and annealed samples. This low temperature growth and higher temperature annealing process enables preservation of smooth films\textemdash as opposed to higher temperature initial growth that leads to much larger surface roughness, as shown for the 20 nm thick film grown at 300 $^{\circ}$C, as discussed in Figure S1 of the Supporting Information (SI). Low surface roughness is critical for subsequent fabrication of on-chip nanophotonic structures such as waveguides and high quality factor optical resonators.

\begin{figure} 
    \includegraphics[width=0.98\textwidth]{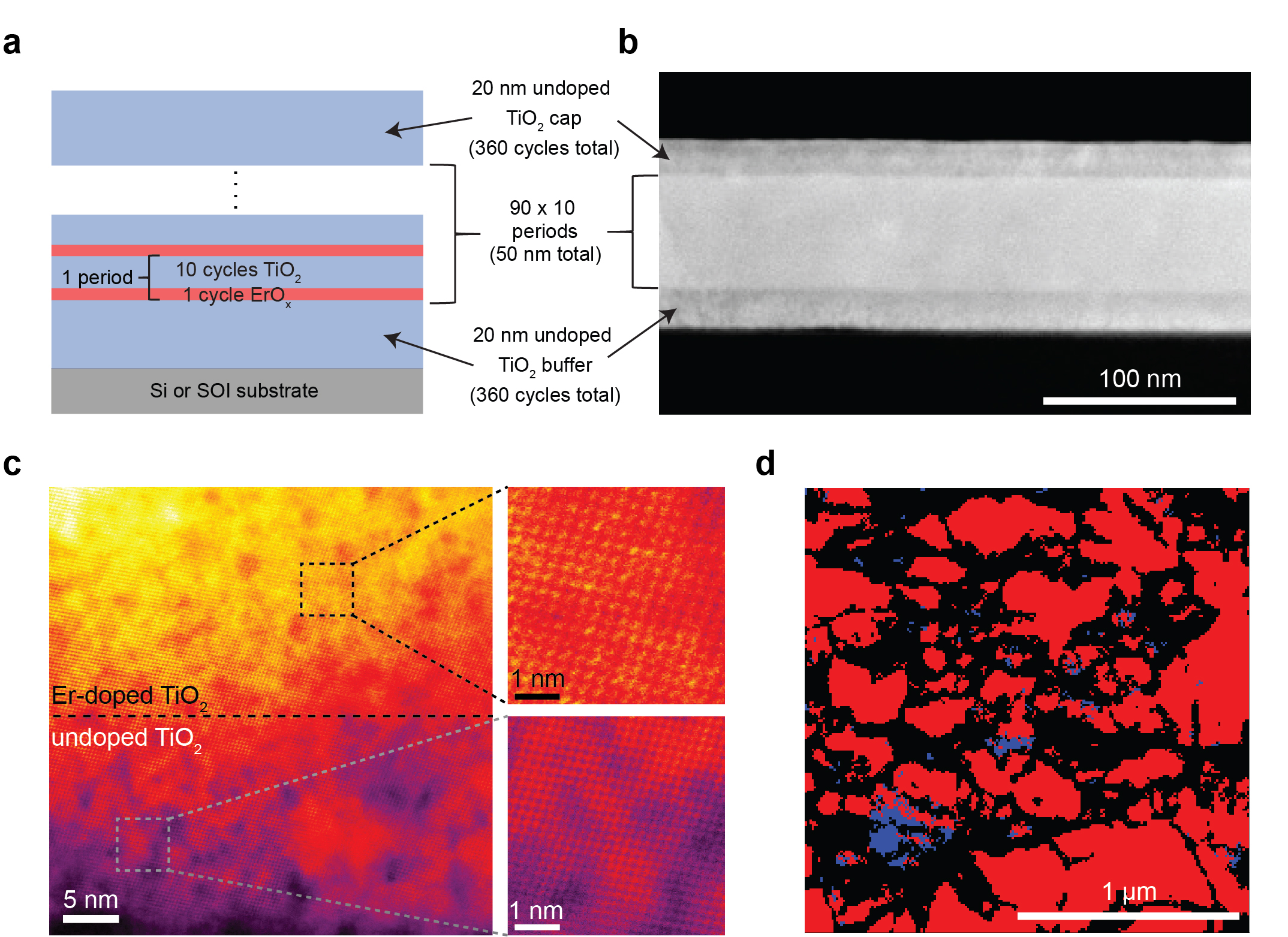}
    \caption{Electron microscopy of Er-doped thin films. (a) Schematic diagram of the `$20/50/20$' doped heterostructure used for SIMS elemental analysis and electron microscopy, consisting of a 20 nm undoped \ce{TiO2} buffer layer, a 50 nm Er-doped \ce{TiO2} layer, and a 20 nm pure \ce{TiO2} capping layer. The thin film is grown on Si(100) or SOI. A thinner `$10/10/10$' heterostructure with the same 1 \ce{ErO_x} to 10 \ce{TiO2} layer ratio is used in the doped region for optical experiments. (b) A low magnification high-angle annular dark-field scanning transmission electron microscopy (HAADF-STEM) image of a sample cross section with an Er doping level of 2950 ppm after oxygen annealing. (c) A high magnification HAADF-STEM image with false color heat map of the same annealed sample as in (b), and focused on the lower boundary between the doped and undoped regions. This cross section clearly shows a single crystalline anatase grain extending through the 20 nm of the undoped \ce{TiO2} near the substrate into the Er-doped region. (d) Electron backscatter diffraction (EBSD) phase map of individual grains within the same `$20/50/20$' film as in (b-c). The red regions consist of individual anatase grains and the smaller and less numerous blue regions are individual rutile grains. The black regions are either amorphous or consist of multiple stacked grains of either phase within the thickness of the 90 nm film.}
    \label{fig:2}
\end{figure}

As described in the previous section, we aim to tailor the Er doping level across a much broader range for various applications by controlling the erbium oxide growth rate. To calibrate the Er concentration, we grew a series of samples by varying the evaporation temperature and pulse time of \ce{Er(thd)_3}, and then analyzed these films using SIMS. The samples prepared for SIMS analysis needed a thicker doped region for sufficient depth resolution so we grew a `$20/50/20$' structure, as illustrated in Figure \ref{fig:2}a (left), comprising a 20 nm nominally undoped \ce{TiO2} buffer layer, a 50 nm layer of Er-doped \ce{TiO2}, and a 20 nm undoped \ce{TiO2} capping layer. The \ce{Er(thd)_3} precursor temperatures, \ce{Er(thd)_3} pulse length, and the experimentally measured Er concentrations via SIMS are listed in Table \ref{Deposition details}. The range of doping provided by these conditions span below 1 parts-per-million (ppm) to a few percent. Previous research has shown that the \ce{Er(thd)_3} vapor pressure can be reduced by approximately 10-fold with a corresponding drop of $\sim25^{\circ}$C in evaporation temperature above 130 $^{\circ}$C. \cite{Sicre1969} Similarly, in this work, the 10-fold reduction of Er dopants between Samples No. 1 and No. 2 in Table \ref{Deposition details} is the result of dropping the evaporation temperature from 160 $^{\circ}$C to 135 $^{\circ}$C. However, the drop in temperature from 135 $^{\circ}$C to 110 $^{\circ}$C dramatically decreased the Er concentration (Sample No. 3 in Table \ref{Deposition details}), where the temperature is out of the reported \ce{Er(thd)_3} evaporation range. Finally, to get sub-ppm level doping we can decrease the pulse length from 8 s to 2 s (Sample No. 4). Regardless, the lower Er doping concentrations (\textless 2 ppm) in Samples No. 3 and No. 4 are significant because they suggest that Er-doping at the level needed for coupling to single ions may be achievable via ALD by reducing the number of \ce{ErO_x} layers overall. 

\begin{table}
    \centering
    \begin{tabular}{c|ccr@{~$\pm$~}l} 
        \hline
        No.\textsuperscript{\emph{a}} & T($^{\circ}$C) \textsuperscript{\emph{b}} & $\Delta t$ (s)\textsuperscript{\emph{c}} & \multicolumn{2}{c}{Doping (ppm)\textsuperscript{\emph{d}}}\\
        \hline
        1 & 160 & 8 & 39200 & 15680 \\
        \hline
        2 & 135 & 8 & 2950  & 885   \\
        \hline
        3 & 110 & 8 & 1.7   & 0.51  \\
        \hline
        4 & 110 & 2 & 0.59  & 0.24  \\
        \hline
    \end{tabular}
    \caption{Sample deposition and Er doping parameters} 
    \label{Deposition details}
    \textsuperscript{\emph{a}} The sample number.\\
    \textsuperscript{\emph{b}} The Er precursor evaporation temperature.\\
    \textsuperscript{\emph{c}} The Er precursor pulse length, same as $\Delta t$ in Figure~\ref{fig:1}a.\\
    \textsuperscript{\emph{d}} The Er concentration determined via SIMS.
\end{table}

To better visualize the crystalline morphology of the Er:\ce{TiO2} thin films we also performed electron microscopy on the doped `$20/50/20$' films used for SIMS analysis (see \textit{Methods}). Specifically, high-angle annular dark-field scanning transmission electron microscopy (HAADF-STEM) imaging was used to provide additional atomic number contrast to better visualize the heavier Er dopants within the \ce{TiO2} matrix (i.e. a brighter pixel corresponds to a higher Z number). Using this technique, we can clearly see in Figure \ref{fig:2}b the increased brightness of the 50 nm thick doped region in a 2950 ppm Er-doped film (same growth conditions as Sample No. 2 of Table \ref{Deposition details}). A corresponding energy-dispersive X-ray spectroscopy (EDS) elemental map of the same cross section in Figure \ref{fig:2}b is shown in the SI (Fig. S2), confirming the presence of Er in the doped region. The as-grown films do not show any regular lattice spacing (see Fig. S3 of SI); however, after the oxygen annealing process, it is common to see large individual anatase grains that are tens of nanometers in thickness extending from the Si substrate\textemdash the crystallinity of which does not appear to be impacted by the onset of the relatively high 2950 ppm Er-doping (Fig. \ref{fig:2}c). The lateral extent of these grains can be visualized with electron backscatter diffraction (EBSD) phase mapping in a scanning electron microscope operating at 10 kV (with an effective probe depth of a few tens of nanometers), as shown in Figure \ref{fig:2}d. The majority of the 4 $\mu$m$^{2}$ area consists of anatase grains ($49.8\%$) while also exhibiting a much lower percentage of smaller rutile grains ($2.0\%$). The exact depth of the grains probed via EBSD varies based upon accelerating voltage and interaction with the Si substrate underneath (see SI Fig.~S3a-b), so it is unclear if the black regions in Figure \ref{fig:2}d merely represent stacked anatase grains with different orientations or pockets of amorphous \ce{TiO2}. Regardless, it is particularly significant that these anatase grains can extend for hundreds of nanometers laterally, as these grains are larger and the sample overall offers higher phase purity than the mixed rutile and anatase grains sizes reported in molecular beam deposited \ce{TiO2} on Si.\cite{Dibos2022} These results suggest that even though our as-grown thin films are amorphous, the oxygen annealing process can induce large anatase grain coarsening, and that these resultant grains might serve as a good local crystal host for erbium ions, all without increasing the top layer surface roughness.

\subsection{Thin Film PLE Measurements}

\begin{figure}
    \includegraphics[width=0.5\textwidth]{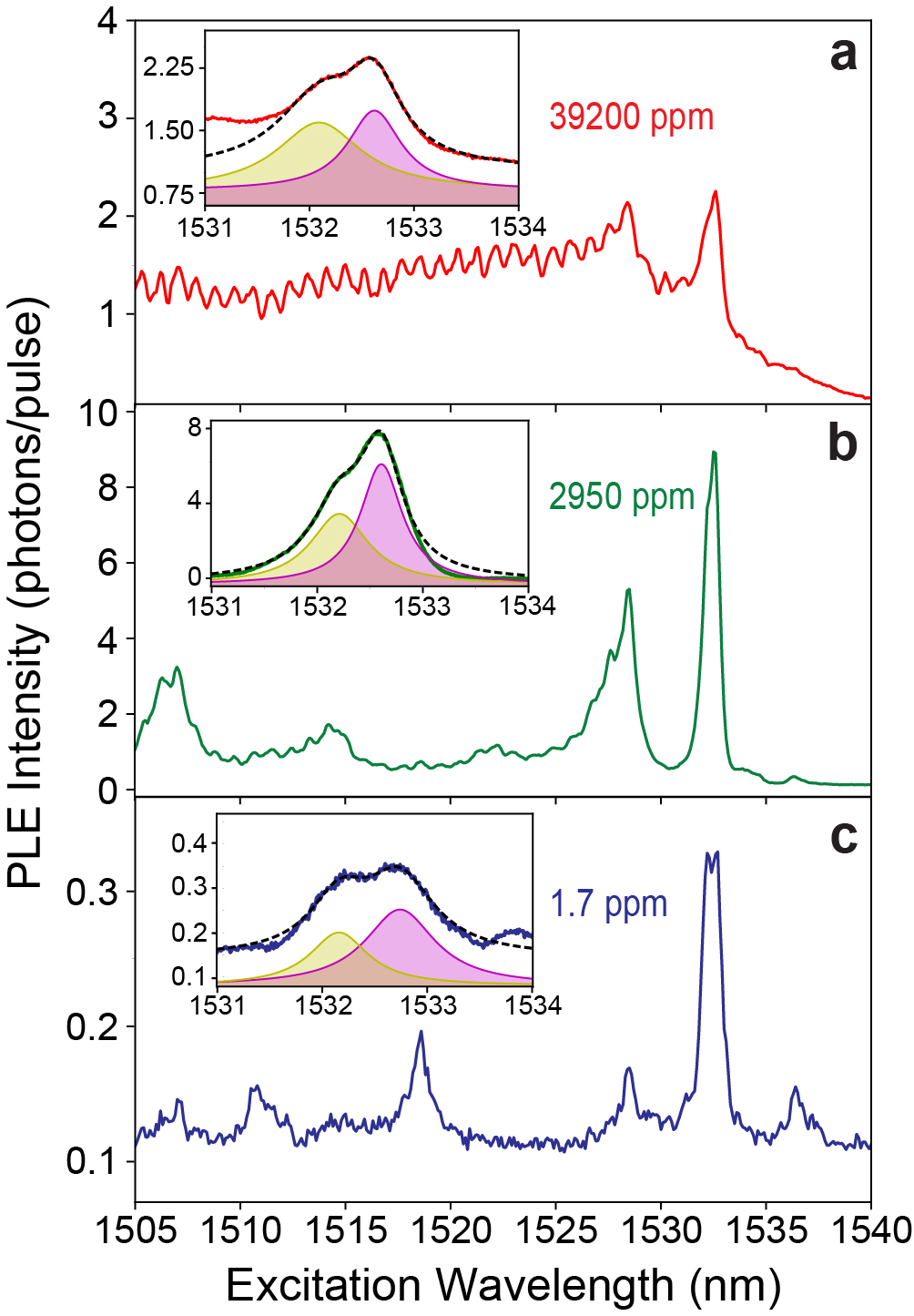}
    \caption{Photoluminescence excitation (PLE) spectra of annealed `$10/10/10$' Er:\ce{TiO2} thin films. In each of the main panels is a broadband PLE scan from 1505 nm to 1540 nm, with a scan step size of 0.1 nm. The thin film Er doping levels for the three samples are (a) 39200 ppm, (b) 2950 ppm, (c) and 1.7 ppm according to Table \ref{Deposition details}. Insets: Corresponding higher resolution PLE scans (step size of 0.01 nm) around the emission peak at 1532.6 nm, giving inhomogeneous linewidths of $79.3\pm2.4$ GHz, $61.4\pm1.7$ GHz, and $102.2\pm5.4$ GHz for the three samples No. 1-3, respectively. Each spectrum is fit to a pair of Lorentzian lineshapes (dashed, black line) with the resultant individual fits highlighted for clarity (purple and yellow, shaded curves).}
    \label{fig:3}
\end{figure}

In order to probe the optical properties of the Er ions within the ALD deposited \ce{TiO2}, we performed photoluminescence excitation (PLE) spectroscopy within the telecom C-band on the aforementioned `$10/10/10$' heterostructures grown on Si substrates, wherein only the central 10 nm layer in the heterostructure contains Er dopants. It is important to point out that for these thinner doped heterostructures, EBSD phase maps suggest that they are dominated by a higher percentage of anatase grains and may extend the full 30 nm thickness of the film (see Fig. S4 of SI). To collect the PLE spectra, the samples were cooled to 3.5 K within a closed-cycle cryostat and excited with a continuously tunable C-band laser using confocal microscopy (details are provided in the \textit{Methods}). As shown in Figure~\ref{fig:3}, we measured PLE spectra for oxygen annealed thin films with the three highest doping levels determined from SIMS (Table \ref{Deposition details}): 39200 ppm, 2950 ppm, and 1.7 ppm. The PLE intensity is the number of detected photons per 3.5 ms collection window after each 1.5 ms laser pulse. It should be noted that for Sample No. 4 at 0.59 ppm the doping level is too low to be measured because of the long optical excited state lifetime of erbium emission, the thin optically active region, and the finite overall collection efficiency of the setup.

All of the as-grown samples had negligible emission in PLE as a result of the underlying disorder in the films. However, after the 30 minute oxygen anneal, all three samples show distinct peaks in the PLE spectra (Fig. \ref{fig:3}). An example of the dramatic increase in PLE brightness for an as-grown versus annealed film is shown in Figure S6 of the SI. Note that the periodic oscillations in the PLE signal are due to Fabry-Perot fringes resulting from partial reflections within the finite thickness of a beamsplitter used in the setup (see SI). The variation in emission intensities among the three samples (Fig. \ref{fig:3}) is mainly attributed to the difference in the doping concentration. In general, higher Er doping typically leads to brighter overall emission; however, as seen by comparing Figures \ref{fig:3}a and \ref{fig:3}b, an order of magnitude increase in doping concentration from 2950 ppm to 39200 ppm does not show a commensurate increase in overall brightness for the emission near 1532 nm. Rather, this reduction in brightness is possibly due to ion-ion quenching and cooperative photon up-conversion, which are common at high Er doping concentrations,\cite{Hehlen97, Vonderhaar21} as the estimated ion-ion spacings at 2950 ppm and 39200 ppm are 1.9 nm and 0.8 nm, respectively. 

For each of the three samples, a higher spectral resolution PLE spectrum (i.e. 0.01 nm step size) near the 1532 nm peak is provided in the inset. Due to the finer spacing, peak fluorescence varies slightly from the corresponding coarse spectrum. Most notable is a pair of closely spaced peaks near 1532.2 nm and 1532.6 nm, similar to what was observed previously in textured growth of polycrystalline \ce{TiO2}.\cite{singh2022} Previous work on Er-doped \ce{TiO2} nanoparticles has experimentally determined the \ce{Y1 \bond{->} Z1} transition to be at 6525 cm$^{-1}$ (1532.57 nm),\cite{Luo2011} similar to the predominant peak at 1532.6 nm in our scans. Future temperature-dependent PL and PLE scans are necessary to confirm this level assignment.\cite{Phenicie2019} However, this value differs from the exact energy of the \ce{Y1 \bond{->} Z1} transition reported for epitaxial anatase Er:\ce{TiO2} growth on \ce{LaAlO3} at 6518.9 cm$^{-1}$ (1534.00 nm).\cite{Shin2022}  As shown for a waveguide-based PLE scan (SI, Fig. S7), we do not see an obvious peak in this range, and the discrepancy could be due to the strain relaxation afforded by the amorphous growth in our polycrystalline anatase films. A deconvolution of these closely spaced peaks enables us to estimate the inhomogeneous linewidth of the 1532.6 nm transition in each of our three samples as $79.3\pm2.4$ GHz, $61.4\pm1.7$ GHz, and $102.2\pm5.4$ GHz, respectively (Figs. \ref{fig:3}a-c, insets). The signal-to-noise ratio for the 1.7 ppm sample is poor (Fig.~\ref{fig:3}, inset), as this sample was pumped with ten times higher laser intensity to get sufficiently large detectable signal above background. Therefore, we suspect that this particular transition is power broadened, and waveguide-based PLE measurements at lower power, suggest an inhomogeneous linewidth closer to 44 GHz for the 1.7 ppm sample (see SI, Fig. S7). In general, the inhomogeneous linewidths of these samples prepared via ALD are broader than epitaxial growth of anatase \ce{TiO2} on \ce{LaAlO3} (12.7 GHz for a 4000 ppm film with similar buffer/capping layer thicknesses)\cite{Shin2022} and molecular-beam deposited growth of highly textured Er-doped anatase \ce{TiO2} on silicon (11.1 GHz with comparable buffer/capping layer thicknesses and doping).\cite{singh2022} However, it is important to note that they are all much narrower than the previously reported results for Er in ALD-grown \ce{Al2O3}, which show broad emission on the scale of THz.\cite{Ronn2016, Ronn2019}
Overall, these findings demonstrate the positive impact of the annealing process in promoting the formation of a more ordered lattice structure and enhancing the optical emission characteristics of Er ions in the anatase \ce{TiO2} matrix.

\subsection{Photonic crystal cavity integration}

\begin{figure}
    \includegraphics[width=\textwidth]{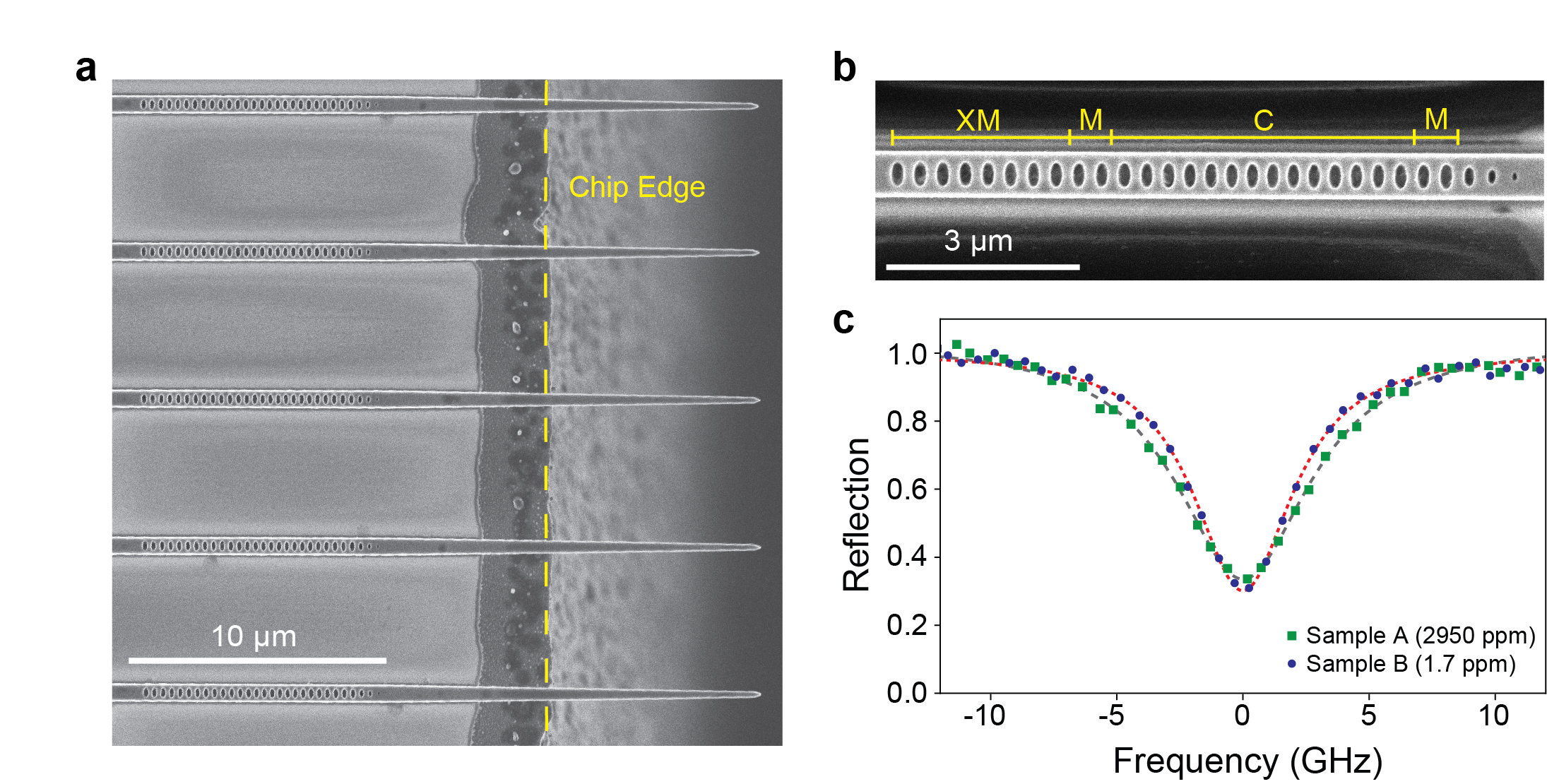}
    \caption{ALD \ce{TiO2}-Si photonic crystal cavities. (a) Top view SEM image of fabricated devices from Sample B, highlighting the inverse tapered waveguide extending off the edge of the SOI chip (to the right of the yellow dashed line). The dark gray region extending 2-3 $\mu$m from the chip edge is due to lateral etching of the original buried oxide layer as part of the undercut process. (b) An SEM image highlighting the nanophotonic cavity composed of elliptically shaped holes etched through the \ce{TiO2} and Si device layers. The 14-hole cavity region (denoted by C) consists of a parabolic taper of the lattice constant, there are two mirror holes (M) on each side of the cavity region, and extra mirror holes (XM) are included on the left-hand side of the device because all measurements are performed in a one-sided coupling configuration via the inverse waveguide taper seen in (a). (c) Normalized laser reflection spectra of the photonic cavities from Samples A (green squares) and B (blue circles) coupled via lensed optical fiber at a temperature of 3.5 K. For these reflection scans, the cavities are tuned onto resonance with the anatase \ce{TiO2} transition of interest ($\lambda = 1532.60$ nm). Using Lorentzian fits, the cavities from Samples A and B have quality factors of Q=$30220\pm890$ (gray dashed line) and  Q=$40680\pm990$ (red dashed line), respectively.}
    \label{fig:4}
\end{figure}

To demonstrate the potential use of these ALD grown Er:\ce{TiO2} films as a platform for telecom on-chip photonic applications, we integrated the same `10/10/10' thin film heterostructure with silicon photonic crystal cavities based on SOI and measured the resultant Purcell enhancement for ensembles in different doping regimes. For this study, we fabricated devices from two samples with different Er concentrations: Sample A (2950 ppm Er, deposition condition No. 2 in Table \ref{Deposition details}) and Sample B (1.7 ppm Er, No. 3 in Table \ref{Deposition details}). The waveguide geometry, photonic crystal cavity, and mirror design lattice parameters are identical to that shown previously\cite{Dibos2022}. In order to compensate for run-to-run etch rate variations, our fabricated devices are designed to have a deliberate elliptical hole size sweep, and there are resonance wavelengths ranging from 1510-1550 nm on each chip. In this work, the desired cavities are nearly resonant at 1532 nm, and as a result the holes are slightly smaller than those used in the original experiments at 1520 nm (rutile phase, \ce{Y1 \bond{->} Z1} transition). In order to improve the one-way coupling efficiency of our silicon nanophotonics, we performed a multi-step process to fully suspend the inverse taper of the waveguide to better mode match to a lensed optical fiber. The substrate and fabrication details are discussed in the \textit{Methods}. To mitigate any potential detrimental defects introduced by the nanofabrication processes, we also oxygen annealed the chips at 400$^{\circ}$C after full device fabrication. SEM images of the final devices are shown in Figure \ref{fig:4}a-b. 

Following the fabrication process, the devices were placed inside a closed-cycle cryostat with a base temperature of 3.5 K for optical characterization using a lensed optical fiber mounted on a 3-axis nanopositioner (see \textit{Methods}). Devices on each sample are screened for cavity resonances 1-2 nm to the blue of the desired optical transition ($\lambda = 1532.6$ nm), after which they are tuned (redshifted) onto resonance via nitrogen gas condensation. Then, each cavity can be characterized by scanning the laser wavelength across the cavity resonance, with their cavity quality factor calculated from the Lorentzian fit of the reflection spectrum as measured with an amplified photodiode. As shown in Figure \ref{fig:4}c, Device A shows a quality factor of Q=$30,220\pm890$ while Device B gives Q=$40,680\pm990$. Overall, the samples etched from Sample A (2950 ppm) had a lower Q than those from Sample B (1.7 ppm) by 30-40\% regardless of detuning from the 1532 nm transition, so this is not due to increased absorption from a higher Er concentration. Rather this uniform discrepancy in Q between the two samples is most likely due to the fact that a more complete \ce{ErO_x} layer (in the case of Sample A) hinders the \ce{Cl2} reactive-ion etching (RIE) step in the doped region of the \ce{TiO2} film, leading to more hardmask erosion during the RIE process and a subsequently lower etch selectivity. Overall, the quality factors of the cavities are comparable to devices fabricated using thinner (21 nm) molecular beam deposited \ce{TiO2} layer with 35 ppm Er doping.\cite{Dibos2022}

\begin{figure}
    \includegraphics[width=\textwidth]{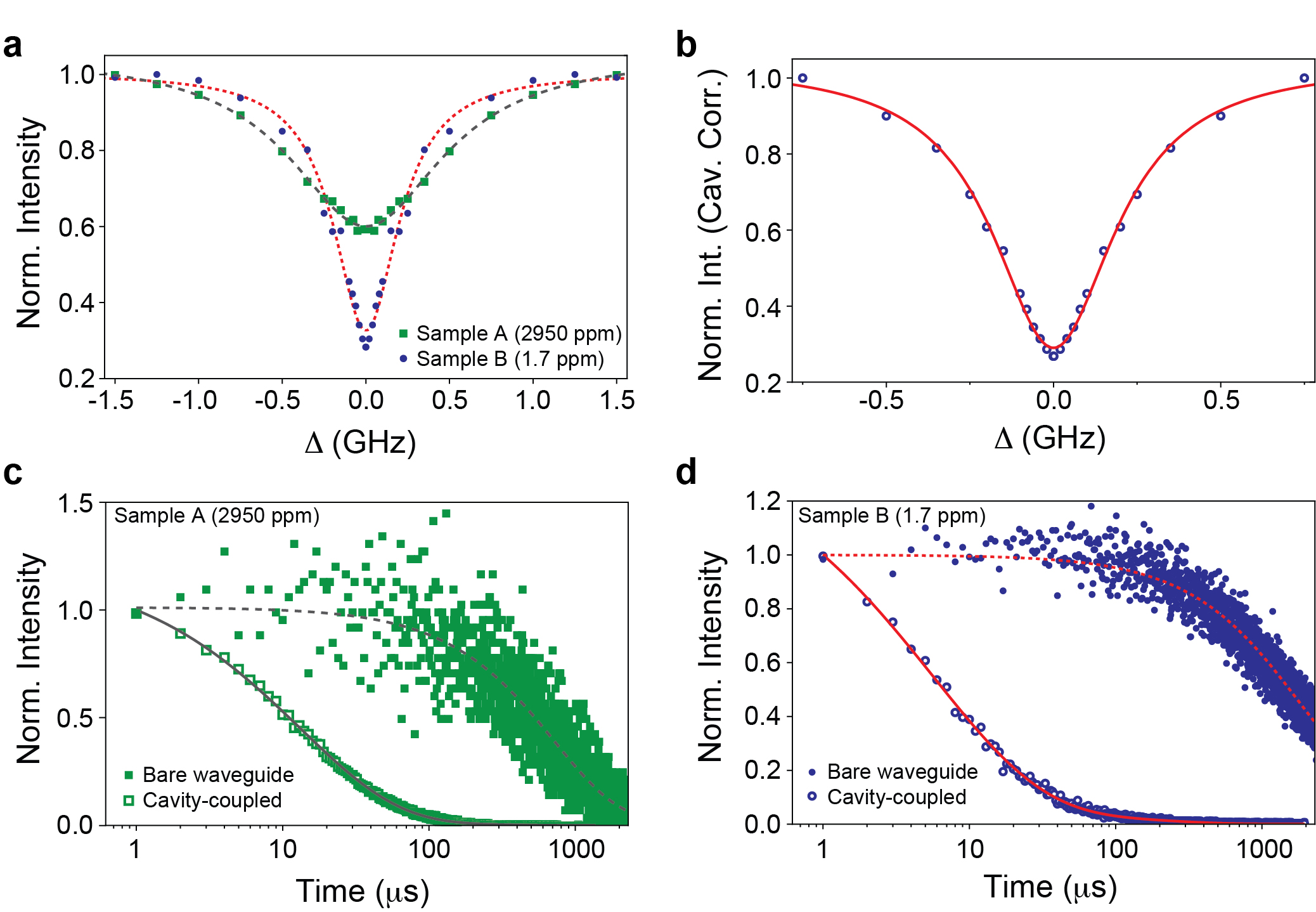}
    \caption{Optical measurements on fabricated devices at T~=~3.5~K. (a) TSHB measurement results performed on bare waveguide devices fabricated from Sample A (green squares) and Sample B (blue circles). The data show the normalized PLE intensity versus the carrier-sideband detuning ($\Delta$) and are symmetric about zero, but they are plotted over the full range for visual clarity. The Lorentzian fit (dashed lines) shows a spectral diffusion HWHM of $0.56\pm0.02$ GHz for Sample A and $0.235\pm0.012$ GHz for Sample B. (b) Similar TSHB measurement results for a cavity-coupled Er ensemble on Sample B, where the normalized intensity has been corrected to account for the lineshape of the same cavity shown in Fig. 4c (blue circles). The Lorentzian fit (solid red line) yields a spectral diffusion linewidth of $0.223\pm0.007$ GHz. (c) Comparison of the normalized ensemble lifetime of Er ions in a bare waveguide device (solid green squares) versus the cavity-coupled device (open green squares) for Sample A (2950 ppm). The bare waveguide lifetime is $725\pm7$ $\mu$s and the cavity-enhanced lifetime is $13.19\pm0.07$ $\mu$s. (d) Comparison of the normalized ensemble lifetime of Er ions in a bare waveguide device (solid blue circles) versus the cavity-coupled device (open blue circles) for Sample B (1.7 ppm). The bare waveguide lifetime is $1718\pm5$ $\mu$s and the cavity-coupled lifetime is $5.7\pm0.2$ $\mu$s. For (c) and (d), the bare waveguide lifetimes are fit to a single exponential and the cavity-enhanced decay times are fit using a stretched exponential as discussed in the text and \textit{Methods}.}
    \label{fig:5}
\end{figure}

The homogeneous linewidth of an optical transition is important for many photonics applications, and this is particularly true for optically addressing single ion quantum memories. A measurement of the homogeneous linewidths of individual emitters would provide the best characterization of disorder within each ion's local environment. However, that measurement is not possible for the high doping concentrations where single emitters cannot be spectrally isolated. Instead, one can perform transient spectral hole burning (TSHB) by sweeping the detuning of two symmetric sideband tones (generated via an electro-optic phase modulator) from the laser carrier frequency to measure the degree of spectral diffusion, which establishes an upper bound on the homogeneous linewidth\cite{Weiss21}. We performed control TSHB measurements via Er embedded in \ce{TiO2} on top of a bare waveguide (no holes). We use 1 ms resonant laser pulses and detect the integrated fluorescence in a 3.95 ms collection window after the end of the pulse as a function of the sideband-carrier detuning frequency ($\Delta$). In Figure \ref{fig:5}a, the TSHB results are shown for Sample A (green squares) and Sample B (blue circles) with the emitters on bare waveguide devices and the half width at half maximum (HWHM) linewidths are $0.56\pm0.02$ GHz and $0.235\pm0.012$ GHz, respectively. The higher doped sample reveals a linewidth that is broader by a factor of two, presumably as a result of the stronger ion-ion interactions. However, in both cases this spectral diffusion linewidth is much narrower than the cavity linewidth, which is $6.47\pm0.19$ GHz and $4.80\pm0.11$ GHz for Samples A and B, respectively. Thus for both samples, coupling between the emitter and cavity lies in the "bad cavity" regime, where the Purcell enhancement of the optical decay rate scales as $Q/V$, where Q is the quality factor and V is the mode volume.

We also performed a TSHB measurement on cavity-coupled Er ensemble for Sample B (Figure \ref{fig:5}b) when the cavity is resonant with the transition at 1532.6 nm. However, for this measurement, the CW laser power is increased by 28$\times$ but the laser pulse lengths are shortened to 2 $\mu$s specifically to address the sub-ensemble of ions that are well-coupled to the cavity. Finally, the normalized intensity as a function of $\Delta$ has been corrected to account for the lineshape of the cavity (shown in Fig. \ref{fig:4}c, green squares), which modifies the sideband-carrier detuning intensity dependence. In this case, the Lorentzian fit (dashed line) yields a spectral diffusion linewidth of $0.223\pm0.007$ GHz, which is comparable to the TSHB linewidth for Sample B ($0.235\pm0.012$ GHz) in Figure \ref{fig:5}a. Both the bare and cavity-coupled spectral diffusion linewidths are comparable to the narrowest value (0.18 GHz) measured previously for molecular beam deposited thin films with thicker buffer layers,\cite{singh2022} suggesting that the quality of the ALD films as a host for Er are comparable to those grown via molecular beam deposition.

The Purcell enhancement of cavity-coupled ions is an increase in the spontaneous emission rate of the emitter due to coupling with the cavity mode. It is a key parameter for enhancing light-matter interaction in quantum and optoelectronic devices. To characterize the Purcell enhancement introduced from the nanophotonic cavities for both Sample A (2950 ppm) and Sample B (1.7 ppm), we performed measurements of the Er emission lifetime using resonant laser pulse excitation. The baseline (non cavity-coupled) optical lifetime was measured from the same control devices as the TSHB measurements, as illustrated by the closed squares (circles) in green (blue) in Figure \ref{fig:5}c (d) for sample A (B). Using a single exponential fit, we found that the natural emission lifetime for the ions on the bare waveguide was approximately 725 $\mu$s for Sample A and 1718 $\mu$s for Sample B. An optical lifetime of 1.7 ms for Sample B is consistent with what is measured in ALD and molecular beam deposited thin films at sub-100 ppm doping levels, suggestive that even after patterning the re-annealed sample does not seem to show signatures of excessive fabrication-induced non-radiative decay. We speculate that the reduced Er optical lifetime for sample A is likely due to ion-ion quenching at higher concentration that serves as a non-radiative pathway.\cite{Michael2008}

We measured the Purcell-enhanced decay rate (open markers) for each sample using the same cavities in Figure \ref{fig:4}c when each cavity is resonant with the Er optical transition at 1532.6 nm. In contrast to the single exponential character of the decay from bare waveguides, the ensembles coupled to each of the cavities show a stretched exponential character as has been seen previously.\cite{Dibos2022, Lei2023} This is due to the varying coupling strength of ions in the ensemble to the cavity at different spatial positions along the cavity and dipole orientation with respect to the TE-like fundamental mode. The time constant for the stretched exponential enhanced lifetime for ions coupled to the resonant cavity was $13.19\pm0.07$ $\mu$s for Sample A and $5.7\pm0.2$ $\mu$s for Sample B, indicating that the best coupled ions in the ensemble have a Purcell factor of $55\pm0.6$ and $301\pm11$, respectively. Overall, we can see that the Purcell factor for Sample A is about 6 times lower than Sample B, and the primary culprits are the reduction in the baseline lifetime (725 $\mu$s versus 1718 $\mu$s) and the modest reduction in cavity Q ($3.02\times 10^4$ versus $4.07\times 10^4$). However, this does not fully account for the difference in the cavity-enhanced decay. Because the Purcell factor measured here comes an ensemble of ions coupled to the cavity, with ions positioned near the cavity nodes being more weakly coupled, the measured stretched exponential decay time only provides a lower bound on the lifetime for the most strongly coupled ions. As a result, it is feasible that the thousand-fold higher concentration of ions in Sample A results in more averaging of the ensemble Purcell factor and hence dilutes the contributions of the best coupled ions to yield a lower enhancement. Overall the Purcell factor measured for Sample B is approximately 50\% better than previously fabricated devices from molecular beam deposited \ce{TiO2} and the Er ions (35 ppm) embedded in the rutile phase grains even though the cavity quality factors were similar.\cite{Dibos2022} Likely this is due to the aforementioned averaging from the higher doping concentration.


\section{Conclusion and Outlook}

Our study showcases a reliable method for growing Er-doped \ce{TiO2} using ALD. One of the most significant findings is the ability to dope these films with Er ions over a large range by controlling the Er-dopant precursor temperature and pulse length. This is a level of doping control not usually provided by ALD and opens up new possible applications at both ends of the doping spectrum (few percent to sub-ppm). With this demonstrable doping range, our ALD process offers potential advantages over other methods, such as molecular beam deposition of doped films and implantation of erbium ions in an undoped films, in terms of translation to industry because of lower cost and increased scalability. Our characterization of the material proprieties after post-growth oxygen annealing revealed the films to be Er-doped anatase phase \ce{TiO2} with a smooth surface, which can be maintained even after the formation of polycrystalline grains that are hundreds of nanometers wide. Furthermore, our optical measurements revealed distinct optical transitions and a substantially narrower homogeneous linewidth upper bound for the characteristic emission peaks around 1532 nm for the annealed films. Additionally, the films with low Er concentration exhibited a natural optical lifetime greater than 1 ms, typical of Er in anatase \ce{TiO2}. These results suggest that Er:\ce{TiO2} thin films grown using ALD are a promising material platform for on-chip quantum memory applications. It is particularly important to note that this critical 400$^\circ$C oxygen annealing process is CMOS-compatible,\cite{Sedky2001} further ensuring industrial scalability. A future area of exploration is the possibility of longer annealing times to further increase anatase grain size or higher annealing temperatures to explore the transition to rutile grains.

Also significant was our coupling of these ALD-grown doped \ce{TiO2} thin films with moderately high quality factor 1D Si photonic crystal cavities that are enabled, in part, because of the low surface roughness after annealing. The resultant Purcell enhancement demonstrates the significant improvement in the spontaneous emission rate of the Er ions due to the evanescent coupling with the fundamental nanophotonic cavity mode. This enhancement plays a crucial role in achieving more efficient and controlled light-matter interactions. In particular, the low-doped sample exhibited a much higher Purcell enhancement in excess of 300, further confirming the potential of these Er:\ce{TiO2} thin films integrated with nanophotonic cavities at dilute concentrations suitable for quantum technologies. 

In order to use these films to spectrally address single Er ion quantum memories, we need to further decrease the doping level to approximately100 ppb,\cite{Dibos2018} depending on the resultant single ion linewidth. It is possible to approach this single ion regime by further reducing the \ce{Er(thd)_3} precursor temperature or restricting the doped \ce{ErO_x} layer to a single cycle\textemdash which would give a nearly 20-fold reduction in doping\textemdash instead of the `$10/10/10$' doped heterostructure employed here. In restricting all dopants to a true delta-doped single layer, it is possible to increase the buffer and capping layer thicknesses to improve the inhomogeneous and spectral diffusion linewidths without needing a thick \ce{TiO2} layer that would inevitably lead to challenges in cavity fabrication and reduced evanescent coupling of ions to the cavity. Finally, the smooth film surface enabled by the low temperature growth and post-anneal can enable facile integration with high quality factor resonators. Particularly interesting is the prospect that the 120$^\circ$C ALD growth process can be performed as a lift-off process with conventional polymer resist windows for even more scalable fabrication processes, rather than top-down etching through the thin films that can potentially damage Er qubits. This work not only highlights that ALD is a promising technique for quantum device fabrication and integration, but could be beneficial for broader Er-based photonics applications, such as on-chip lasing and amplifiers.


\section{Methods} \label{Methods}

\subsection{Er-doped \ce{TiO2} thin film ALD growth}
In this study, we used a Fiji G2 plasma-enhanced ALD (PEALD) system, which can be operated with and without plasma, to produce the Er-doped \ce{TiO2} thin films. The deposition process involved the use of titanium isopropoxide (TTIP) (from Sigma-Aldrich with 5N purity), heated to 68 $^{\circ}$C, and water (\ce{H2O}), maintained at ambient conditions, as precursors for the thermal deposition of \ce{TiO2}. During the \ce{TiO2} deposition, the Ar flow was set to 110 sccm to purge out the by-product and excessive precursors from the chamber. For the Er doping, the Tris(2,2,6,6-tetramethyl-3,5-heptanedionato)erbium (\ce{Er(thd)_3}) is from Strem Chemicals with 3N purity and it can react via oxygen plasma to form atomic layers of erbium oxides. The operating power of the oxygen plasma was 300 W with \ce{O2} flow of 25 sccm and Ar flow of 55 sccm. Unless noted otherwise, all of the thin films are deposited on lightly boron-doped standard silicon substrates from NOVA Electronic Materials and Silicon Valley Microelectronics (SVM). The as-grown thin film thicknesses are routinely characterized by X-ray reflection (XRR) and this is used to estimate the growth rate of \ce{TiO2} under these conditions to be 0.5-0.6 \AA~per cycle. The XRD and XRR measurements were completed with a Rigaku SmartLab located in the University of Chicago MRSEC. After the deposition, thermal annealing was conducted at 400 $^{\circ}$C for 30 min a rapid thermal annealing system (Annealsys) with a consistent \ce{O2} flow rate of 500 sccm and resultant pressure near 700 Torr. AFM scans on the as-grown and post-annealed thin films were conducted with a Bruker FastScan Atomic Force Microscope (AFM) with ScanAsyst. 


\subsection{Electron Microscopy}
To prepare samples for transmission electron microscopy (TEM), we used a Zeiss NVision SEM-FIB system for specimen `lift-out'.  Prior to the lift-out process, a $2\,\mu\text{m}$ thick carbon layer was deposited on the sample to prevent Ga ion beam damage. A $20\times2\times5\, \mu\text{m}^{3}$ ($\text{length}\,\times\,\text{width}\,\times\,\text{height}$) specimen was milled from the sample with its height along the (001) direction of the silicon substrate (perpendicular to the wafer surface) and transferred onto a TEM grid with an Omni-probe. Following the lift-out procedure, the specimen was thinned further via focused ion beam to $100$ nm. Finally, the specimen was double-side-polished in a Gatan PIPS II precision ion polishing system with $0.3\,\text{kV Ar}^{+}$ to remove any amorphous residue left by the FIB milling process. The polished specimens then were loaded into a Thermofisher Spectra 200 aberration corrected STEM to obtain atomic resolution images of the Er-doped \ce{TiO2} thin film. The images were taken with 200 kV electron accelerating voltage yielding a typical resolution of $\sim 70$ pm. To enhance the contrast of Er atoms, a collection angle larger than 75 mrad is used for the high annular dark-filed (HAADF) imaging. The EBSD grain mapping was performed with JEOL JSM-IT800 SEM with a backscattering electron detector, with the probe current fixed at $4.7$ nA for $10$ kV accelerating voltage and $7.1$ nA for $20$ kV.

\subsection{Thin film PLE measurements}

To characterize the emission properties of the Er ions doped in the \ce{TiO2}, we probed the Er ion optical properties using photoluminescence excitation (PLE) spectroscopy near 1.5 $\mu$m. The samples were cooled to 3.5 K within a Montana s50 Cryostation. The measurement configuration employed a Toptica CTL 1500 tunable laser passed through a pair of fiber-coupled acousto-optic modulators (MT80-IIR30-Fio-PM0.5-J1-A-Ic2, AA Opto Electronic) then passed into free space, where they are routed by a 50:50 beamsplitter into a 0.65 NA objective lens focused on the thin film sample using a 3-axis periscope stage (Newport). Emitted photons are routed back through the beamsplitter and into the fiber network via collimator after two long-pass filters (FELH1500, Thorlabs) for detection. Single telecom photons were detected via superconducting nanowire single-photon detector (SNSPD) from Quantum Opus within a second cryostat, which was optimized for near-1.5 $\mu$m detection near 80\% quantum efficiency. The collected light is gated by an additional modulator (AMM-55-4-50-1550-2FP, Brimrose) to prevent latching of the SNSPD during laser excitation. Event timing and single photon counting are handled by a dedicated Pulse Streamer 8/2 (Swabian Instruments) and Time Tagger Ultra (Swabian Instruments), respectively. For each PLE scan, at each wavelength, one measurement shot consisted of an excitation pulse of 1.5 ms followed by a collection interval of 3.5 ms, averaged over 10,000 shots for each wavelength, to generate a histogram of the single photon counts detected during the collection window.
The overall intensity of the collected signal is integrated for each wavelength to produce a PLE spectrum. A diagram of the thin film PLE measurement configuration is presented in Figure S5a of the SI.

\subsection{Nanophotonic device fabrication}

The silicon-on-insulator wafers used for nanocavity fabrication have a 220 nm thick Si device layer, 2 $\mu$m thick buried oxide (Soitec, Inc). The device layer is lightly boron doped with a resistivity of 10~$\Omega \cdot$cm. Following ALD thin film growth, all fabrication is performed in the Center for Nanoscale Materials cleanroom at Argonne National Laboratory and is similar to the process flow outlined in the supplementary information of prior work.\cite{Dibos2022} As part of the device fabrication flow, we need to strip the \ce{SiO2} hardmask in buffered oxide etch (BOE) in a later step, so prior to the \ce{SiO2} hardmask deposition we performed a 30-minute oxygen anneal (Annealsys) at 400 \textdegree C in pure oxygen, which makes the Er:\ce{TiO2} film highly resistant to later BOE exposure. We use standard electron beam lithography (JEOL 8100) with ZEP 520A resist on top a hardmask of \ce{SiO2} deposited via plasma enhanced chemical vapor deposition (PECVD, Oxford PlasmaLab 100). Fluorine-based etching is used for mask transfer to the \ce{SiO2}, \ce{Cl2} is used to etch through the \ce{TiO2} layer, and HBr/\ce{O2} is used to etch through the Si device layer. All etching is performed in an Oxford PlasmaLab 100 inductively coupled plasma (ICP) reactive ion etcher (RIE). As mentioned in the main text, the etch rate of the Er-doped \ce{TiO2} decreases with increased doping concentration. A thick reference sample of commensurate doping is used to calibrate the Cl$_2$ etch rate of \ce{TiO2} for the different doping concentrations. Unlike the waveguide geometry employed previously~\cite{Dibos2022}, the inverse tapered waveguides are fixed length, designed to be truncated, masked, and vapor etched for better coupling between the on-chip waveguide and the lensed fiber. The full details of the taper design, hardmask chemistry, undercut process, and fiber-to-waveguide coupling efficiency estimates will be discussed in a forthcoming manuscript. Within our large array of devices there are clusters of nominally identically designed cavities with the same hole size (the hole size is swept to account for run-to-run variations in the \ce{TiO2} and Si etch rates. The ideal device resonances are approximately 1541-1543 nm at room temperature to account for the 12 nm resonance blueshift when cooled to 3.5 K. Experimentally the cavity Qs at room temperature are within about 10\% of each other for a particular etching run/sample, and each resonance is typically within $\sim$2 nm of the other devices within the cluster.

\subsection{Device Measurements}

We measured the photonic crystal cavity devices in a separate Montana Instruments s100 Cryostation with a sample base temperature of 3.5 K. The nanopositioner, lensed fiber configuration, nitrogen gas tuning, optical modulators, optical fiber network, and single photon detectors are nearly identical to work published previously.\cite{Dibos2022} The schematic of the optical measurement setup for nanocavities can be seen in Figure S5b of the SI. For both the transient spectral hole burning (TSHB) and lifetime control measurements (Fig. \ref{fig:5}a,c-d), we used bare waveguide devices fabricated on the same chip as the cavity devices. For these control measurements, one shot consisted of a 1 ms laser pulse lengths and 3.95 ms photon collection interval during which the single photon counts were detected via superconducting nanowire single photon detector (SNSPD) in a separate cryostat. For the TSHB control measurements the total number of detected photons was integrated for each sideband detuning frequency ($\delta$) output with a function generator to the phase modulator. For the bare waveguide lifetime control measurements (Fig. \ref{fig:5}c-d, filled markers), there was no sideband tone and the lifetime is fit to a single exponential of the single photon detection times (via a quTAG time tagger) after accounting for the SNSPD dark count rate (30 Hz). For the cavity-coupled transient spectral hole burning measurement (Fig. \ref{fig:5}b) the input CW laser power was increased by 28$\times$ but the laser pulse lengths were shortened to 2 $\mu$s to probe spectral diffusion for the ions that are well-coupled to the cavity and will have cavity-enhanced decay rates. The normalized intensity as a function of $\Delta$ (Fig. \ref{fig:5}b) has been corrected to account for the Lorentzian lineshape of the cavity (FWHM=4.8 GHz, shown in Fig. \ref{fig:4}c green dots), which modifies the intensity dependence on sideband-laser detuning. Similar to the cavity-coupled TSHB measurements, the cavity-enhanced lifetime (Purcell) measurements are performed using 2 $\mu$s resonant laser pulses but with no driven sideband tone. For all device measurements there is a 2 $\mu$s delay between the end of the incident laser pulse and the opening of the collection path intensity modulator to the superconducting nanowire single photon detector. For the cavity-coupled TSHB and lifetime measurements the cavity resonance is systematically redshifted onto the optical resonance at 1532.6 nm using nitrogen gas ice condensation, while periodically monitoring the spectral position of the cavity via a CW laser reflection scan. For the cavity-coupled lifetime measurements shown in Figure \ref{fig:5}c-d (open markers), the solid line fits are in the form of a fast-decay stretched exponential and a slow single exponential as described in detail in the SI of previous work,\cite{Dibos2022} where we use a non-linear model fit to find the stretched exponential time constant ($\tau_{fast}$), and we interpret 1/$\tau_{fast}$ as the fastest Purcell-enhanced lifetime of the ensemble.

\begin{acknowledgement}

The authors would like to thank D. Czaplewski, C. S. Miller, and R. Divan for assistance with device fabrication. The authors acknowledge the Q-NEXT Quantum Center, a U.S. Department of Energy, Office of Science, National Quantum Information Science Research Center, under Award Number DE-FOA-0002253 for support (C. J., G. D. G., C. P. H., I. M., S. K. S, D. D. A., S. G., and A. M. D.). AFM characterization, oxygen annealing, electron microscopy, and cavity device fabrication were performed at the Center for Nanoscale Materials, a U.S. Department of Energy Office of Science User Facility, supported by the U.S. DOE, Office of Basic Energy Sciences, under Contract No. DE-AC02-06CH11357. Additional support (M. H. and J. W.) for TEM characterization is supported by Quantum Information Science research funding from the U.S. DOE, Office of Science User Facility. Additional materials characterization support (M. T. S., S. E. S., and F. J. H.) was provided by the U.S. Department of Energy, Office of Science; Basic Energy Sciences, Materials Sciences, and Engineering Division. Additional support for cryogenic and optical measurements was provided by the U.S. Department of Energy, Office of Science; Basic Energy Sciences, Materials Sciences, and Engineering Division. Additional support (M. K. S.) for growth capabilities was provided by the Center for Novel Pathways to Quantum Coherence in Materials, an Energy Frontier Research Center funded by the U.S. Department of Energy, Office of Science, Basic Energy Sciences under Award No. DE-AC02-05CH11231.

\end{acknowledgement}

\begin{suppinfo}

The Supporting Information contains additional experimental and analysis details, such as: AFM scan of an ALD \ce{TiO2} film grown at 300 $^{\circ}$C, TEM EDS elemental mapping of the doped `$20/50/20$' heterostructure, comparison of crystallinity of Er-doped \ce{TiO2} pre- and post- 400 $^{\circ}$C oxygen annealing, EBSD phase maps for thinner `$10/10/10$' \ce{TiO2} films used for optical measurements, experimental configurations used for optical measurements, PLE results from as-grown samples, and PLE measurements on a 1.7 ppm waveguide-based device.

\end{suppinfo}

\bibliography{achemso-demo}

\end{document}